\documentclass{article}

\usepackage{amssymb}
\usepackage[english]{babel}
\usepackage{graphicx}
\usepackage{graphics}
\usepackage{lineno}

\begin{document}

\title{Deviance Information Criteria for Model
Selection in Approximate Bayesian
Computation}

\author{Olivier Fran\c cois$^1$ \and Guillaume Laval$^2$}
\maketitle

\vspace{1cm}

\noindent{$^1$ Universit\'e Joseph Fourier Grenoble, Centre National de la Recherche Scientifique, TIMC-IMAG UMR 5525, 38042 Grenoble, France.}

\noindent{$^2$ Institut Pasteur, Human Evolutionary Genetics, Paris, France.}

\begin{abstract}
Approximate Bayesian computation (ABC) is a class of algorithmic methods in Bayesian
inference using statistical summaries and computer simulations. ABC has become popular in
evolutionary genetics and in other branches of biology. However model selection under ABC algorithms
has been a subject of intense debate during the recent years. Here we propose novel
approaches to model selection based on posterior predictive distributions and approximations of
the deviance. We argue that this framework can settle some contradictions between the computation
of model probabilities and posterior predictive checks using ABC posterior distributions. A
simulation study and an analysis of a resequencing data set of human DNA show that the deviance
criteria lead to sensible results in a number of model choice problems of interest to population
geneticists.
\end{abstract}

\pagestyle{empty}

\section*{Introduction}

Approximate Bayesian Computation (ABC) is a class of Monte-Carlo algorithms for parameter inference based on summary statistics instead of the full data (Beaumont et al 2002, Marjoram et al 2006, Beaumont 2010). More specifically, ABC algorithms use simulations from a stochastic model to generate random samples from an approximation of the posterior distribution of a multidimensional parameter, $\theta$, after reduction of the original data, $y_0$, into a set of summary statistics, $s_0 = s(y_0)$. Here we will consider that $s_0$ are the only data available to ABC analyses, and will refer to (Robert et al 2011) for  discussions related to the sufficiency of the summary statistics. ABC methods found their origin in evolutionary genetics (Pritchard et al 1999, Tavar\'e 2004), where they have been fruitfully applied to the inference of demographic history of several species (Lopes and Beaumont 2009, Csill\'ery et al 2010, Beaumont 2010). Examples of analyses encompass the evaluation of alternative scenarios of human evolution (Fagundes et al 2007, Patin et al 2009, Laval et al 2010), inference in demographic models of population expansion, bottleneck or migration (Thornton et al 2006, Fran\c cois et al 2008), population structure and adaptation (Bazin et al 2010).

Current ABC algorithms fall into broad subclasses of methods that extend the standard subclasses of computational algorithms used in Bayesian statistics. The first class of algorithms makes use of the rejection algorithm to accept parameters generating simulated data close to the observations (Pritchard et al 1999). The rejection algorithm performs the following steps: 1) Generate a candidate value $\theta$ from a prior distribution; 2) Simulate a data set, $y$, from a generating mechanism using the parameter $\theta$, and compute the set of summary statistics ${\bf s} = s(y)$; 3) Accept the value of $\theta$ if the (Euclidean) distance between ${\bf s}$ and $s_0$ is less than $\epsilon$, a prespecified error value; 4) If rejected, go to 1). For this basic algorithm, the accepted values $(\theta_i)$ form a random sample from an approximation of the posterior distribution.

The above approximation becomes exact as $\epsilon$ goes to zero, but  the algorithm is then highly inefficient. Recent techniques improve the approximation of the posterior distribution by applying linear or non-linear transforms (Beaumont et al 2002, Wegmann et al 2009, Leuenberger and Wegmann 2010, Blum and Fran\c cois 2010). In those improvements, the accepted values of the parameter, $\theta_i$, are weighted by a quantity that depends on the distance between ${\bf s}_i$ and $s_0$. Then they are adjusted according to a regression transform, for example, $ \theta_i^* = \theta_i - b^T ( {\bf s}_i - s_0)$, where $b$ is a vector of linear regression coefficients (Beaumont et al 2002). Several studies have provided evidence that the transformed parameters form a significantly better approximation of the posterior distribution than the non-transformed ones (Beaumont et al 2002, Blum and Fran\c cois 2010), and regression adjustments are now widely used by ABC practioners (Thornton 2009, Cornuet et al 2009, Lopes and Beaumont 2009). Two other classes of algorithms implement Markov chain Monte Carlo methods without likelihood (Marjoram et al 2003, Bortot et al 2007) and iterative algorithms that were originally inspired by sequential Monte Carlo samplers (Sisson et al 2007, Beaumont et al 2009, Toni et al 2009).

An important aspect of ABC is its use for model selection in addition to parameter estimation. In general the aim of model selection is to find models receiving the highest posterior probabilities among a finite subset of candidates. Bayesian statisticians have devised numerous ways to evaluate and select models for inference (Gelman et al 2004). Assuming that there are $M$ models under consideration, the Bayesian paradigm includes model selection in the inference step, taking the model label as an additional parameter, $m$. In decision theoretic approaches, model choice is performed on the basis of posterior probabilities, $p(m | s_0)$, which are proportional to the marginal probabilities, $p(s_0 | m)$. In ABC these probabilities can be crudely estimated by counting simulations from model $m$ that fall at a distance less than a fixed value to the observed data. More sophisticated estimators of posterior model probabilities can be found in (Beaumont 2008) or in (Leuenberger and Wegmann 2009). Alternatively sequential Monte-Carlo algorithms can also used to estimate model probabilities via iterated importance sampling procedures (Toni and Stumpf 2010).

Model selection using ABC algorithms has been recently questioned (Templeton 2009, Beaumont et al 2010, Csill\'ery et al 2010, Robert et al 2011). Here we point out a potentially serious concern when selecting models on the basis of approximate posterior model probabilities. Because approximate model probability estimates are based on the rejection algorithm and ignore regression adjustments on parameter samples, we argue that model choice based on these probabilities does not apply to the (approximate) models in which we eventually make inference. To see this, assume $\theta|m = \theta_m$, and let 
\begin{equation} 
p_{\epsilon}(\theta_m |s_0) \propto {\rm Pr}( \| {\bf s} - s_0 \| \leq \epsilon | \theta, m) p(\theta_m) \,  ,
\end{equation}
be the approximation of the posterior distribution obtained from the rejection algorithm, where $p(\theta_m)$ denotes the prior distribution on the parameter $\theta_m$ for model $m$. The joint distribution defining model $m$ is then equal to
\begin{equation} 
p_{\epsilon}(\theta_m , s_0 | m) = p_{\epsilon}(\theta_m | s_0 ) p_{\epsilon}(s_0 | m) \,  .
\label{eq:m}
\end{equation}
Regression adjustments replace $p_{\epsilon}(\theta_m | s_0 )$ with another distribution $p_{\rm reg}(\theta_m | s_0 )$, which is generally closer to the exact posterior distribution. Clearly this change modifies the joint distribution in equation (\ref{eq:m}). Thus a model chosen on the basis of $p_{\epsilon}( m | s_0 )$ can be different from the model in which we eventually estimate parameter uncertainty.

In the next section, we define two information theoretic criteria for model selection based on measures of model fit penalized by an estimate of the model complexity. While our focus in on regression methods, the ideas introduced in the present study apply to any ABC algorithm.
The approach shares similarities with the popular Akaike information criterion (AIC, Akaike 1974) which is valid for the comparison of nested models (Burnham et al 2002, Johnson and Omland 2004, Ripley 2004, Carsten et al 2009). The assumption of nested models is seldom appropriate to ABC, and we develop a statistical theory of approximate deviance information criteria (DIC), a generalization of AIC that does not require the assumption of nested models (Spiegelhalter et al 2002, Gelman et al 2004). Then we provide an example of ABC analysis where model choice based on approximate probabilities disagree with the prediction of adjusted models and DICs. Using simulations, we study the relevance of the proposed information criteria to inference in population genetics under various models of demographic history and population structure. In the last part of the study, we present an application to an empirical genetic data set of 20 noncoding DNA regions resequenced from 213 humans (Laval et al 2010), and we use DICs to question the replacement of Neanderthals by modern humans.

\section*{Theory}

In this section we describe model selection criteria based on posterior predictive distributions and approximations of the deviance.

\paragraph{Information theoretic criteria}

In Bayesian analyses, the deviance information criterion summarizes the {\it fit} of a model by the posterior expectation of the deviance, $\bar{D}$, and the {\it complexity} of a model by its {\it effective number of parameters}, $p_{D}$ (Spiegelhalter et al 2002). The models that receive the highest support from the data are those with the lowest values of the DIC. More specifically, the definition of DIC is
\begin{equation}
{\rm DIC} = \bar{D} + p_{D},  
\end{equation}
where the deviance is minus twice the logarithm of the likelihood, $D(\theta) = - 2 \log p(s_0|\theta)$,  $\bar{D}$ is the expected deviance
\begin{equation}
 \bar{D}  =  {\rm E}_{\theta | s_0} [ D(\theta)]  \, , 
\end{equation}
and $p_D$ is the difference between $\bar{D}$ and the deviance evaluated at a particular point estimate, $D(\hat{\theta})$. An example of $\hat{\theta}$ often used in applications is the estimate of the posterior mean of the model parameter.

A complication arises when models are defined hierarchically. In hierarchical models there is a hidden parameter $\varphi$, and the posterior distribution decomposes as 
follows
\begin{equation}
p(\theta, \varphi | s_0) \propto p(s_0| \varphi ) p(\varphi|\theta) p(\theta) \, . 
\end{equation}
In this situation, several definitions of the deviance  and DIC have been proposed depending on the {\it focus} of the model (Spiegelhalter et al 2002; Celeux et al 2006). For example focusing on $\theta$, the deviance can be taken equal to 
\begin{equation}
D(\theta) =  -2 \log \left( \int_{\varphi} p(s_0 | \varphi ) p(\varphi|\theta) d\varphi \right) \, , 
\label{eq:phi}
\end{equation}
and the computation of DIC should be modified accordingly.

\paragraph{A hierarchical model approach to ABC}

Without regression adjustments, one way to define ABC is as a hierarchical Bayesian model in which the simulated summary statistics are viewed as latent variables.
In this hierarchical model, the posterior distribution decomposes as 
\begin{equation}
p(\theta, {\bf s} | s_0) \propto p(s_0|{\bf s}) p({\bf s}|\theta) p(\theta)
\end{equation}
where ${\bf s}$ are the simulated statistics, and $\theta$ becomes the ``hyper-prior'' parameter. 

In this model, $p({\bf s}|\theta)$ is the probability of the generating the summary statistics, ${\bf s}$, with parameter $\theta$. To make use of a hierarchical framework, we define a surrogate likelihood function 
\begin{equation}
p( s_0 | {\bf s} ) \equiv K_{\epsilon} (  s_0 - {\bf s}  ) 
= \frac1{\epsilon} K \left( \frac{ s_0 - {\bf s}   )}{\epsilon} \right) \, , 
\end{equation}
where $K$ is a density function, called the {\it kernel}, and $\epsilon$ is the error parameter. The distribution $p( s_0 | {\bf s} )$ can be viewed as a model for the observation error, and ABC performs exact inference under the assumption of model error (Wilkinson 2008). The hierarchical model  reformulation of ABC dates to the work of Marjoram et al (2003) who used it as a rationale for defining MCMC algorithms without likelihood. This point of view has also proven useful in a variety of theoretical works on ABC (Bortot et al 2007, Ratmann et al 2009, Wilkinson 2008). Regarding the basic rejection algorithm, the definition amounts to choosing a uniform density function over the interval $(0,1)$ for the  kernel. In this case, we obtain  
\begin{equation}
p( s_0 | \theta) \approx \int_{\bf s} p( s_0 | {\bf s} ) p( {\bf s} | \theta) d{\bf s} = {\rm Pr}( \| s_0 - {\bf s}  \| < \epsilon  | \theta ) \, . 
\end{equation}

Extensions of the rejection algorithm use non-uniform kernels. For example, Beaumont et al (2002) implemented the Epanetchnikov function which is popular in density estimation. Because we want to relate the quantity $\log K_{\epsilon}(  s_0 - {\bf s}   )$ to a natural measure of model fit, we take the Gaussian kernel 
\begin{equation}
 K( u ) =  \frac1{\sqrt{2\pi}} e^{-u^2/2} \;, \quad u \in \mathbb{R}.
\end{equation}
With this choice, the quantity $-2 \log K_{1}(  s_0 - {\bf s}  )$ has a natural interpretation as the sum of squares error between observed and simulated statistics.

The surrogate model presented above has a two-level hierarchy. Following Spiegelhalter et al (2002) or Celeux et al (2006), distinct definitions of DIC can be proposed, depending on whether the focus is on the fit of the summary statistics to the observed ones or on the model parameters themselves. Focusing on the parameter level allows us to better evaluate the predictive power of the fitted models, and we next introduce two definitions for an approximate deviance at this level. 

A first way to define a Bayesian deviance is by considering a posterior predictive average of a ``low level'' deviance 
\begin{equation}
{\rm dev}(\theta) = -2 \,  {\rm E}_{{\bf s}| \theta} [ \log p(s_0 | {\bf s})  ] =  -2 \int_{{\bf s}}  \log p(s_0| {\bf s})  p( {\bf s}  | \theta) d{\bf s} \, . 
\end{equation}
In this case, the expected Bayesian deviance is 
$$
\bar{D}^1 = {\rm E}_{\theta|s_0} [{\rm dev}(\theta)] = -2 \, {\rm E}_{{\bf s}| s_0 } [ \log p(s_0 | {\bf s})  ] 
$$ 
With this definition, a Monte-Carlo estimate of the expected deviance can be easily computed from the simulated data as follows
\begin{equation}
\bar{D}^1 \approx  
-  \frac2n \sum_{j = 1}^n \log \left( K_{\epsilon}( {\bf s}^j - s_0  ) 
\right)  \, ,
\end{equation}
where the ${\bf s}^j$ are summary statistics obtained from the posterior predictive distribution $p({\bf s} | s_0)$. To compute the penalty $p_{D}^1$, we generate $n$ summary statistics ${\bf s}^j$ from $p( {\bf s} |\hat{\theta})$, where $\hat{\theta}$ is a point estimate of $\theta$, for example an estimate of the posterior mean, E[$\theta|s_0 $]. Applying the same formula as above, we come with an estimate $\bar{D}^1(\hat \theta)$ that we use to define $p^1_D$
\begin{equation}
p_D^1  = \bar{D}^1  - \bar{D}^1(\hat \theta) \, . 
\end{equation}

Though the focus is on the parameter $\theta$, the previous definition of a deviance is not equivalent to equation (\ref{eq:phi}). A definition of the deviance in a hierarchical model consistent with this equation is as follows
\begin{equation}
D(\theta) = -2 \log p(s_0 | \theta) = -2 \log \left( \int_{{\bf s}} p(s_0 | {\bf s}) p({\bf s} | \theta) d{\bf s} \right) ,
\end{equation}
which is also equal to
\begin{equation}
D(\theta) = -2 \log \left( E_{{\bf s}| \theta}[ K_{\epsilon}( {\bf s} - s_0 )  ] \right) .
\end{equation}
With this definition, an estimate of the expected deviance requires two levels of Monte Carlo integration
\begin{equation}
\bar{D}^2 =  E_{\theta|s_0} [D(\theta)]
\approx  - \frac2m \sum_{i = 1}^m \log  
\left( 
  \frac1n \sum_{j = 1}^n K_{\epsilon}(  {\bf s}^j_i - s_0  ) 
\right) 
\end{equation}
where we have $m$ replicates, $(\theta_i)_{i=1,\dots,m}$, from the approximate posterior distribution, $p(\theta | s_0)$, and each ${\bf s}^j_i$  is sampled from $p({\bf s} | \theta_i)$, $j = 1, \dots, n$. To compute $p_{D}^2$, we generate $n$ summary statistics, ${\bf s}^j$, from the conditional distribution $p(s|\hat{\theta})$, where $\hat{\theta}$ is a point estimate of $\theta$, and we set 
\begin{equation}
\bar{D}^2(\hat \theta) 
\approx  - 2 \log  
\left( 
  \frac1n \sum_{j = 1}^n K_{\epsilon}( {\bf s}^j - s_0  ) 
\right) \, . 
\end{equation}
Then we define
\begin{equation}
p_D^2  = \bar{D}^2  - \bar{D}^2(\hat \theta) \, . 
\end{equation}
Both definitions of ${\bar D}$ and $p_D$ lead to distinct definitions of an information criterion, DIC$_i = \bar{D}^i + p_{D}^i$, $i = 1,2$.  DIC$_2$ has the advantage of defining DIC for ABC models more rigorously than DIC$_1$, but it has the disadvantage of being computationally more intensive. 

Unlike model probabilities, $\bar{D}$ and $p_{D}$ can be computed from any approximation of the posterior distribution. Using linear or non-linear regression adjusments, we can consider the transformed parameter posterior distribution, $p_{\rm reg}(\theta_m |s_0)$, instead of $p_{\epsilon}(\theta_m |s_0)$. To compute DIC, we then replace the $\theta_i$'s by their adjusted values $\theta_i^*$'s, sampled from the modified posterior distribution, and generate posterior predictive densities from these values. In the sequel, DIC$_i$ will refer to predictive distributions generated from adjusted parameters.

To motivate the use of information criteria and illustrate some of the issues presented in the introduction, we consider an example where samples of size $n = 20$ are simulated from a Gaussian disribution of mean $\mu_0 = 2$ and standard deviation $\sigma_0 = 3$ (then assumed to be unknown). The data are summarized by their empirical mean, standard deviation, skewness and kurtosis, and the sample size is known. 

We observed $s_0 = (2.00, 3.11, -0.78, 0.14)$. For these data two models are hypothesized. The sampling distribution of first model is a Gaussian distribution where the parameter, $\theta = (\mu, \sigma^2)$, corresponds to the mean and the variance. The prior distribution on $\mu$ is a Gaussian distribution of mean 2 and standard deviation $10$. The prior distribution on $\sigma$ is an inverse-exponential distribution of rate 1. The  sampling distribution of the second model is a Laplace distribution of mean 3 and rate $\lambda$. The prior distribution on $\lambda$ is an exponential distribution of rate 1. 

To perform ABC analyses, we simulated 10,000 samples from each model, and pooled the 20,000 vectors of statistics into a single data set. Using an acceptance rate of 10$\%$, we estimated model probabilities using the R package {\tt abc} (R core team 2010). This package computes the proportion of accepted simulations under both models, and also implements the weighted logistic regression method of (Beaumont 2008). 
The Bayes factor, defined as the ratio of marginal probabilities for two models $m_1$ and $m_2$ can be estimated as the ratio of counts in favor of $m_1$ and $m_2$ (Pritchard et al 1999, Grelaud et al 2009). 
The proportion of accepted simulations from model 2 (Laplace) was 0.83. Assuming a uniform prior distribution on models 1 and 2 we obtained an approximation of the Bayes factor equal to BF $\approx 5.02$. The logistic regression estimate for the posterior probability of model 2 was 0.85, and we obtained BF $\approx 5.75$. According to Jeffrey's scale on Bayes factors (Jeffreys 1961), there would be substantial evidence in favor of the Laplace model over the Gaussian model. 

In a second stage, we performed regression adjustments to the approximate posterior samples. The observed value of the kurtosis statistic was outside the tails of the posterior predictive distribution under the Laplace model. In contrast it was within the tails of the posterior predictive distribution under the Gaussian model. Thus there is an apparent contradiction between the computation of model probabilities and the predictions from the posterior distribution. To better understand this contradiction, we drew the exact posterior distribution of $\sigma^2$ under the Gaussian model (an inverse-Gamma$(11, 1+ 9.5 v_{0}^2)$ distribution, where $v_{0}^2$ is the empirical variance). Although posterior density approximations are improved by the adjustment method (Figure 1), the estimates of model probabilities did not account for such improvements. Then we computed DICs for the Gaussian and Laplace models. Under the Gaussian model, we obtained DIC$_1 = 4.5$ and DIC$_2 = 3.2$. Under the Laplace model, we obtained DIC$_1 = 10.1$ and DIC$_2 = 4.5$. Once the corrections were applied, DIC indicated that the Gaussian model was a better choice than the Laplace model. 

\begin{figure}
\begin{tabular}{c}
\includegraphics[width=12cm,angle=0]{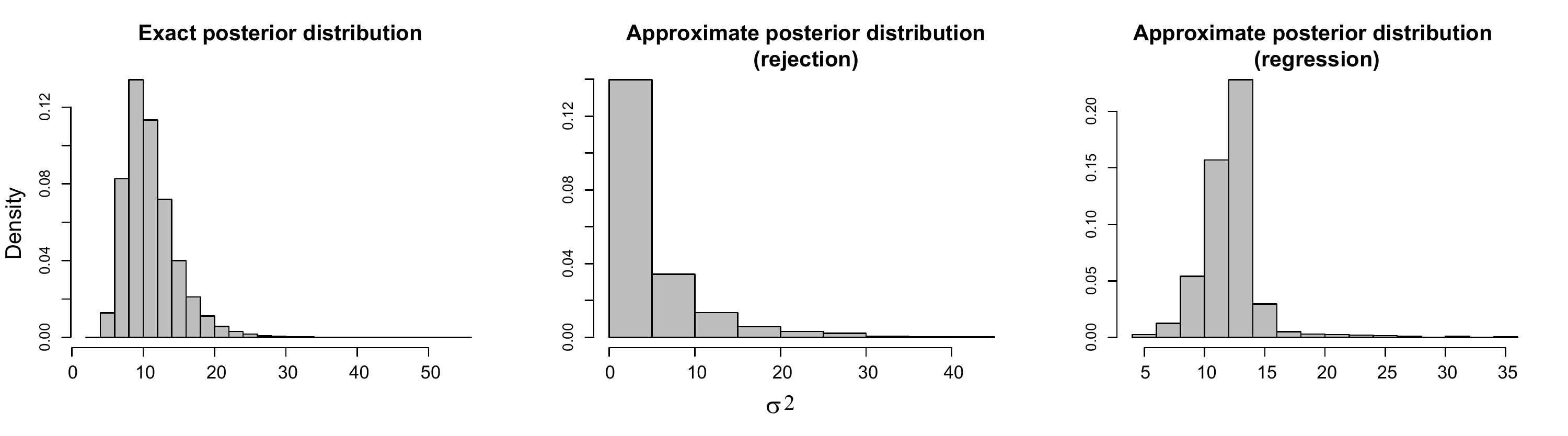}
\end{tabular}
\caption{\em Posterior distributions of parameter $\sigma^2$ (Gaussian model). A) The exact posterior distribution. B) The approximation obtained from the rejection algorithm. C) The approximation obtained with the regression adjustement.}
\end{figure}

To investigate whether our analysis was robust, we replicated it 100 times with values of $s_0$ sampled from the same Gaussian distribution in each replicate. The average value of the Bayes factor was around 3.45 (3.58 when using the logistic regression method) in favor of the Laplace model which obtained the highest posterior probability in 100$\%$ of the replicates. In contrast, the Gaussian model was preferred in 100$\%$ of the replicates when DIC was used. The mean value of the information criterion was around 4.1 (2.4 for DIC$_2$) under the  Gaussian model, whereas is was around 12.4 (3.7 for DIC$_2$) under the Laplace model.

\section*{Population genetic models}

For each simulated data set and each model considered afterwards, we performed 10,000 simulations under specified prior distributions (100,000 for the human data). We ran ABC analyses using an acceptance rate of 10$\%$ ($1\%$ for the human data). To compute DIC$_1$ we additionally created $n = $1,000 replicates from the posterior predictive distribution. For DIC$_2$, we used $m= 200$ and $n= 200$ replicates.

\paragraph{Demographic models.}

We generated coalescent simulations under selectively neutral models of micro-evolution for 3 distinct demographic scenarios: a sudden decline in population size (bottleneck without recovery), a constant population size, an exponentially growing population size. In these data, fifty diploid individuals were genotyped at 20 non recombining loci. The data were simulated as DNA sequences under an infinitely many-sites model using the computer program {\tt ms} (Hudson 2002). 

For each of the three models, we simulated one hundred replicates of the data with fixed parameter values. In {\tt ms}, the parameters are expressed in units of the current population size, $N_0$. In our simulations, the normalized mutation rate was equal to $\theta = 2N_0 \mu = 3$. In the bottleneck model, the population size shrunk to a fraction $x = 1/4$ of its ancestral value, and this event occurred $t = 0.2 N_0$ generations in the past. In the expanding population model, the expansion rate was set to $\alpha  = 2$. 

The prior distribution on the mutation rate, $\theta$, was uniformly distributed over $(0,15)$ for all models. In the bottleneck model, the date of the bottleneck event (in unit of $N_0$) was uniformly distributed over $(0,1)$, and $\log_{10} x$ was uniformly distributed over $(0,1.5)$. In the expansion model $\log_{10} \alpha$ was also uniformly distributed over $(0,1.5)$. ABC analyses were performed using the following summary statistics: the Tajima's estimator $\pi$, computed as the mean number of differences between pairs of sequences, Tajima's $D$ (Tajima 1989), and Fay and Wu's $H$ statistic (Fay and Wu 2000). The three statistics were averaged over the 20 loci.

We applied ABC analyses and deviance information criteria to population genetic data simulated under a bottleneck, a constant population size and an expansion model (100 data sets for each demographic model). Table 1 reports congruent results for  DIC$_1$ and DIC$_2$, and Figure 2 reports the outcome of model selection for the three models.

\begin{table}
\begin{center}
\begin{tabular}{ccccc}
\hline
& & & & \\
&     & Bottleneck & Constant size & Expansion \\
     & & & &\\
\hline
& & & &\\

Bottleneck$\star$ &  & 13.66 (7.72) & 12.83 (7.99) & 18.36 (9.82) \\
&  & 10.08 (8.80) & 10.59 (6.08) & 17.04 (9.71) \\

& & & &\\

\hline
& & & &\\

Constant size & & 40.06 (10.93) & 2.83 (0.18) & 3.15 (0.30)\\
              & & 30.24 (14.71) & 2.71 (0.17) & 2.95 (0.39)\\

& & & &\\

\hline
& & & &\\

Expansion     & & 17.48 (4.59) & 3.65 (0.64) & 3.36 (0.23)\\
              & & 11.40 (4.12) & 3.58 (0.69) & 2.87 (0.17)\\

& & & &\\
\hline
\end{tabular}
\end{center}
\caption{\em Comparisons of demographic models using DIC$_1$ and DIC$_2$. The rows correspond to the models used for simulating the data, and the columns correspond to the models under which inference was performed. The values represent the mean and standard deviation of DICs computed over 100 independent replicates for each model. $\star$: Values under the bottleneck model were computed with the median of the posterior deviance instead of their mean, because the median is less sensitive to large deviations.}
\end{table}

Although DIC$_1$ is generally greater than DIC$_2$, the two measures produced highly correlated results (Figure 3). In these examples, both criteria agreed in their evaluation of models. When the data were simulated under the bottleneck model, reported estimates were obtained with the median of the posterior deviance instead of their mean (The mean provided highly variable results). The preferred model was the bottleneck model in 61/100 replicates. When the data were simulated under the constant population size model, the preferred model was the constant size model in 81/100 replicates. When the data were simulated under the expanding population size model, the preferred model was the expanding population model in 89/100 replicates. Overall, models that did not generate the data were selected in a small but non-neglectible number of replicates (23$\%$). For data simulated under the bottleneck model, the lower performances can be explained by the relatively recent date of bottleneck event and by the choice of diffuse prior distributions. Both factors contribute to the difficulty of distinguishing between a bottleneck and a constant size model. In addition there is great variability in the simulated summary statistics, and this can explain why the bottleneck and constant population size models were difficult to tease apart.

\begin{figure}
\begin{center}
\includegraphics[width=12cm,angle=0]{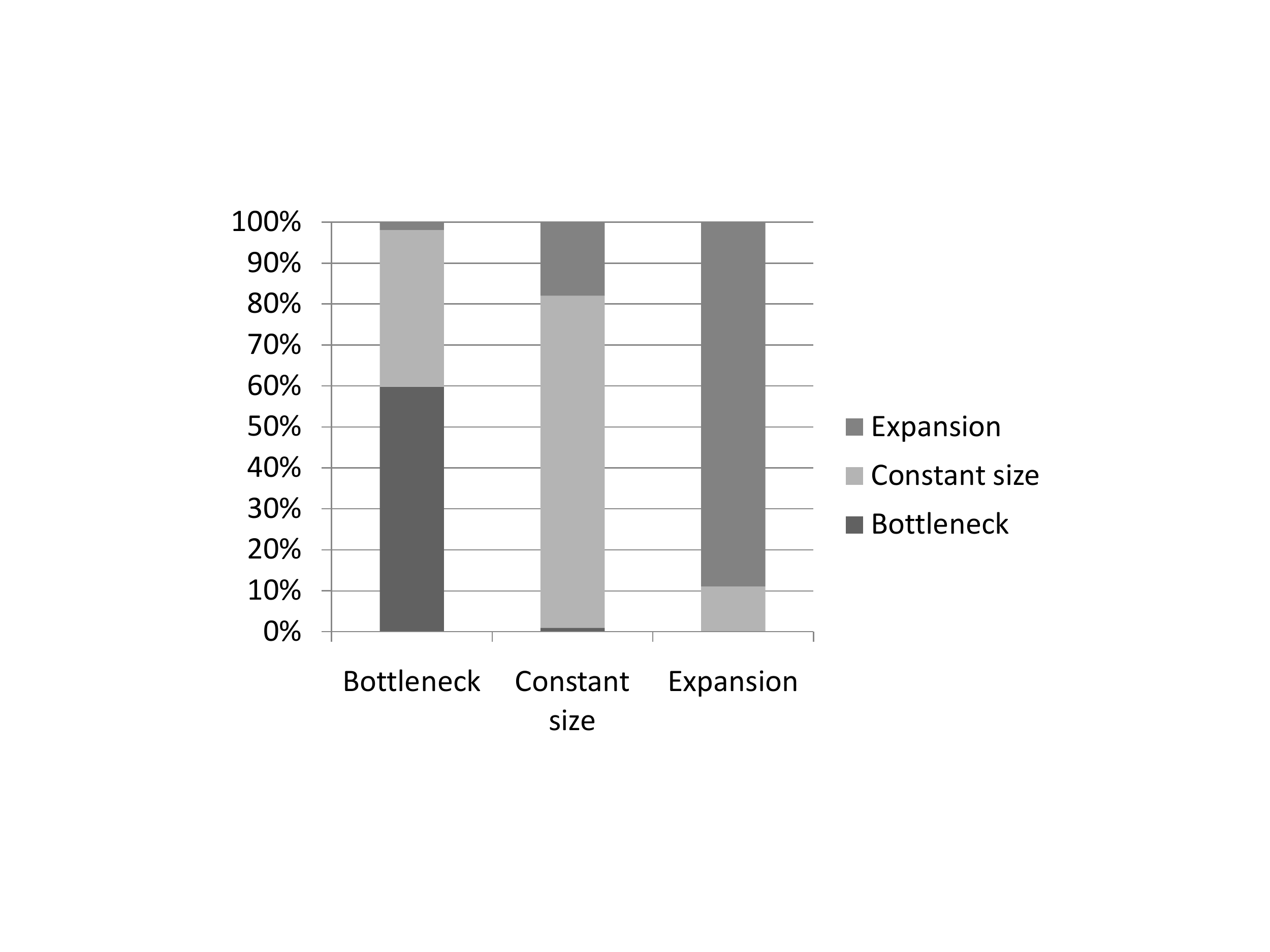}
\end{center}
\caption{\em Demographic scenarios. Model choice using the deviance information criterion, DIC$_2$, when a specific scenario is assumed. See the text for the definition of models. The frequencies were computed over 100 independent replicates from each model. The same results were obtained with DIC$_1$.}
\end{figure}

\begin{figure}
\begin{center}
\includegraphics[width=10cm, angle=0]{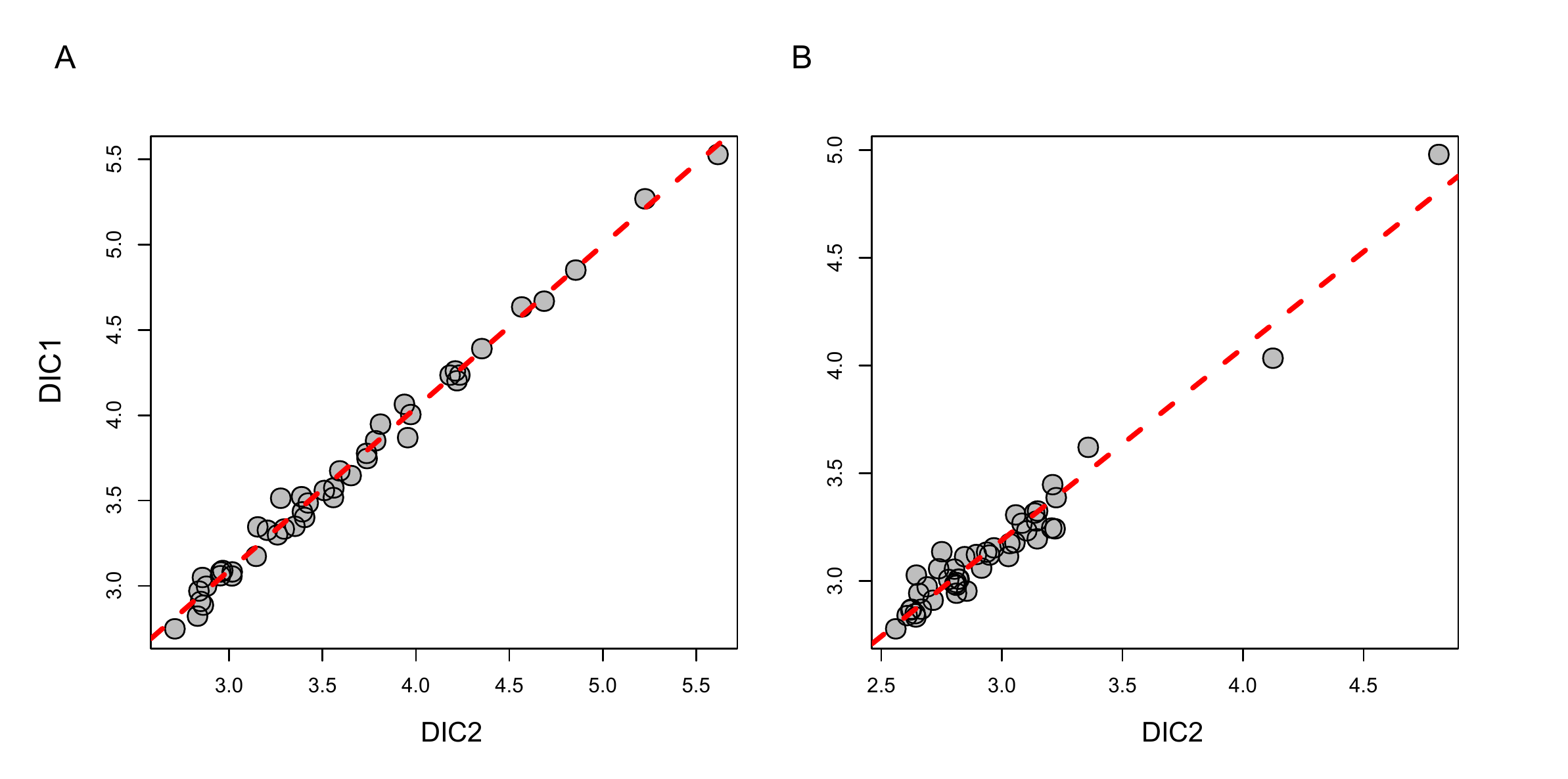}
\end{center}
\caption{\em Correlation between DIC$_1$ and DIC$_2$. A) Inference under a population expansion model for data simulated under a constant population size model. B) Inference under a constant population size model for data simulated under an expanding population model.}
\end{figure}

To gain insight on model choice, we examined the results obtained for a particular replicate in further details. The replicate was obtained from the bottleneck model, where the effective mutation rate, the time of the bottleneck event and the severity of the bottleneck $1/x$ were set to the values (3, .2, 4). The observed summary statistics were equal to $\pi = 8.58$, $D = 1.32$ and $H = 9.42$.  Under the bottleneck model, the point estimates of the effective mutation rate (posterior mean = 2.90, .95 CI = [1.34, 4.36]), the time of the bottleneck event (posterior mean = 0.35, .95 CI = [0.04, 0.62]) and the severity of the bottleneck (posterior mean = 5.01, .95 CI = [2.18, 47.81]) were close to the values used in the simulation.  DIC$_2$ was equal to 14.13 (median estimate), whereas it was equal to 17.83 under the constant population size model. In this case the DIC slightly favored the model that generated the data. Under the bottleneck model, the observed values of the summary statistics were indeed within the tails of the posterior predictive distributions whereas they were outside the tails in the constant size population model (Figure 4). We also investigated why in some case DIC failed to select the model from which the data were simulated. For one bottleneck data set, the observed summary statistics were equal to $\pi = 6.77$, $D = 0.56$ and $\log H = 1.71$. DIC$_2$ was equal to 10.11 (median estimate) under the bottleneck model, whereas it was equal to 6.08 under the constant population size model. Under both models, the observed values of the summary statistics were within the tails of the posterior predictive distributions, but the posterior distributions had larger tails under the bottleneck model than under the constant size model, and DIC favored the most parcimonious model (constant population size). This shows that in some cases the data do not contain enough information to discriminate between models, and highlights the difficulty of making model inference under coalescent models.

\begin{figure}
\begin{center}
\includegraphics[width=8cm, height = 8cm, angle=0]{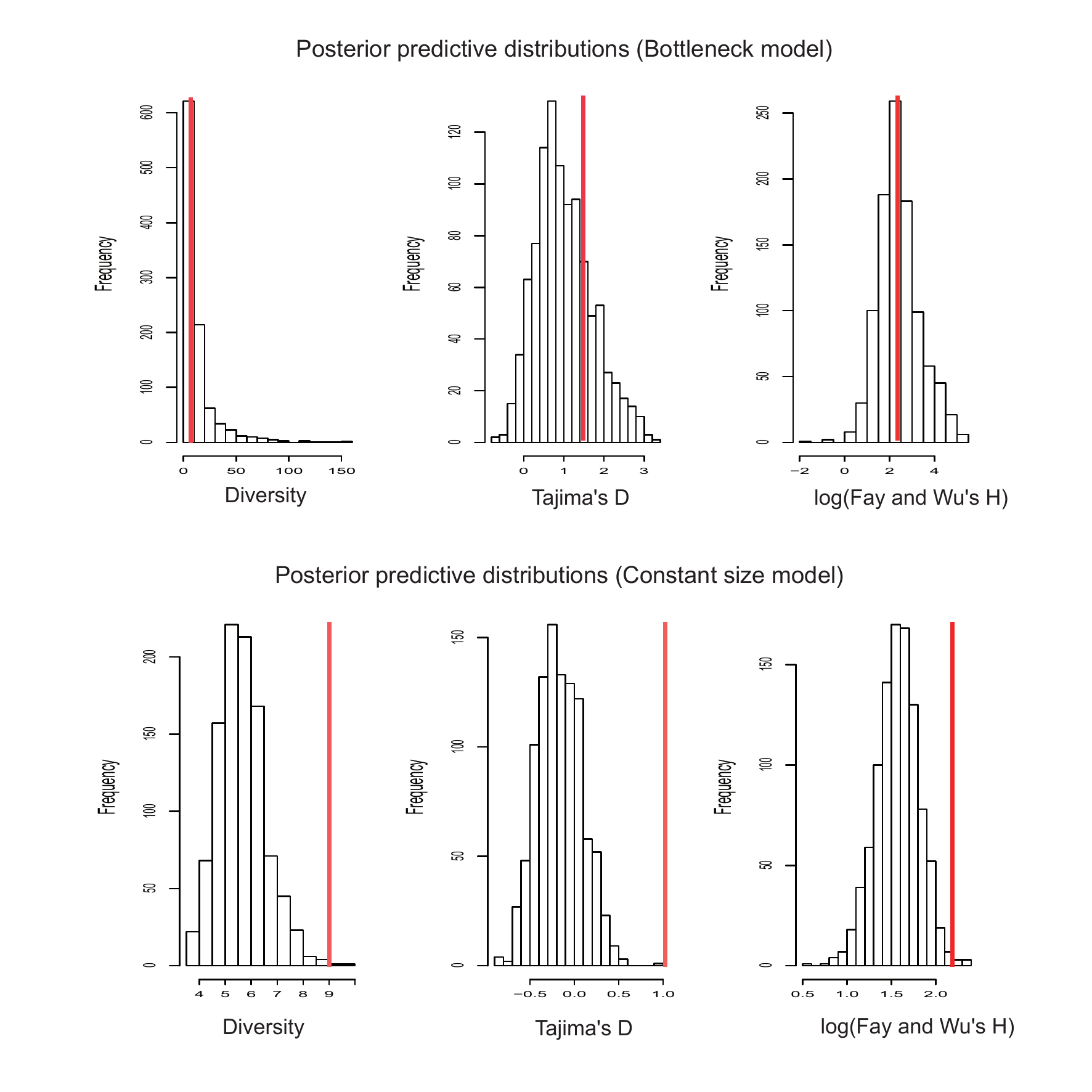}
\end{center}
\caption{\em Posterior predictive distributions of $\pi$, Tajima's $D$ and the logarithm of Fay and Wu's $H$. The observed summary statistics were simulated under a bottleneck model, and were equal to $\pi = 9.77$, $D = 1.40$ and $H = 9.98$. Inferences were performed under a bottleneck model and a constant population size model. The vertical bars correspond to the observed summary statistics.}
\end{figure}

\paragraph{Isolation with migration.}

Next we generated coalescent simulations under 3 distinct models of population structure: divergence of two subpopulations, divergence with unidirectional gene flow from subpopulation 1 to subpopulation 2, and divergence with migration between two subpopulations.  Fifty diploid individuals in each population were genotyped at 100 non recombining loci. For each of the three models, we simulated one hundred replicates of the data with fixed parameter values. For these models, the parameters are expressed in units of the current subpopulation size, $N_0$. In our simulations, the normalized mutation rate was equal to $\theta = 2N_0 \mu = 4$. In each model, the two subpopulations had equal population size (both were equal to $N_0$). For the divergence model, the split occurred $0.7 N_0$ generations ago. In the divergence with migration model, the normalized migration rates were equal to $m_1N_0 = 40$ and $m_2N_0 = 30$, whereas in the unidirectional gene flow model, we took  $m_1N_0 = 40$ and $m_2N_0 = 0$.  

When running ABC inferences, subpopulations sizes were parameterized as $\nu_1 N_0$ and $\nu_2 N_0$ respectively, and we used uniform prior distributions on $(0,3)$ both for $\nu_1$ and $\nu_2$. In all models, the divergence time was uniformly distributed over $(0,N_0)$. Normalized migration rates were uniformly distributed over $(0, 100)$. In addition to the three statistics $\pi$, $D$ and $H$ used in demographic models, we also computed an $F_{\rm ST}$ statistic for each data set according to Hudson's estimator (Hudson et al 1992). The statistics were averaged over the 100 loci.

We applied ABC analyses to 300 data sets simulated under the above models of population structure and gene flow between two subpopulations. Table 2 reports congruent results for DIC$_1$ and DIC$_2$, and Figure 5 describes the results of model selection for the three models of population structure.

\begin{table}
\begin{center}
\begin{tabular}{ccccc}
\hline
& & & & \\
&  & Isolation & Asymetric Migration & Migration \\
 & & & &\\
\hline
& & & &\\

Isolation &  & 2.97 (.15) & 3.37 (.25) & 4.67 (.71) \\
          &  & 2.67 (.12) & 3.18 (.39) & 4.15 (.81) \\

& & & &\\

\hline
& & & &\\

Asymetric   & & 3.13 (.25) & 2.79 (.19) & 4.10 (.31)\\
Migration   & & 2.84 (.23) & 2.65 (.20) & 3.41 (.33)\\

& & & &\\

\hline
& & & &\\

Migration     & & 3.02 (.12) & 3.02 (.20) & 3.82 (.25)\\
              & & 2.73 (.11) & 2.85 (.19) & 3.25 (.20)\\

& & & &\\
\hline
\end{tabular}
\end{center}
\caption{\em Comparaison of isolation with migration models. The rows correspond to the models used for simulating the data, and the columns correspond to the models under which inference was performed. The values represent the mean and standard deviation of DIC$_1$ and DIC$_2$ computed over 100 independent replicates for each model.}
\end{table}

As for the simulations of demographic scenarios, DIC$_1$ and DIC$_2$ produced highly correlated results, and agreed in their evaluation of models. When the data were simulated under the divergence model, the preferred model was the divergence model in 98/100 replicates. When the data were simulated under a model with unidirectional migration, the preferred model was the model with asymetric migration in 92/100 replicates. When the data were simulated under the migration model, DIC favored either a divergence model (49/100 replicates) or a model with asymetric migration (51/100), but it never chose the model model that generated the data. As previously, DICs favored explanatory models that required the smallest number of parameters. A typical example of posterior predictive checks for data simulated under the bi-directional migration model is displayed in Figure 6. For the 3 models, the observed summary statistics were between the tails of their posterior predictive distributions. The summary statistics used in this example were not informative enough to distinguish between gene flow and divergence (Nielsen and Wakeley 2001, Hey and Machado 2003). The fact that DIC did not select the model that generated the data should not be considered as an error. The correct interpretation is that the three models were equally good at reproducing the observed statistics, and DIC values confirmed that we could not make any strong decision in this case.

\begin{figure}
\begin{center}
\includegraphics[width=12cm,angle=0]{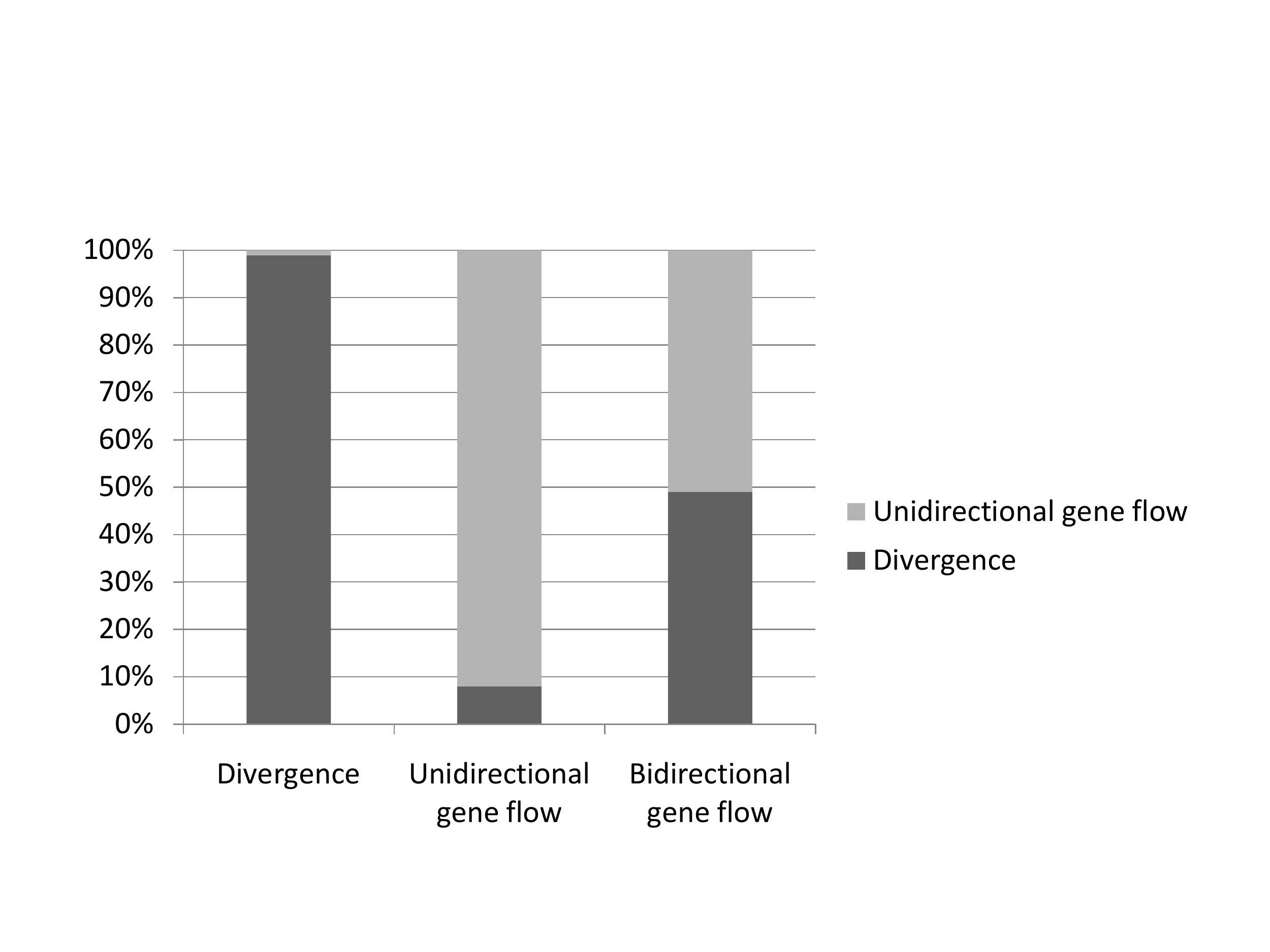}
\end{center}
\caption{\em Population structure and migration scenarios (100 replicates). Model choice using the deviance information criterion, DIC$_1$, when a particular scenario is assumed. See text for the definition of models. The frequencies were computed over 100 independent replicates from each model.}
\end{figure}

\begin{figure}
\begin{center}
\includegraphics[width=12cm,angle=0]{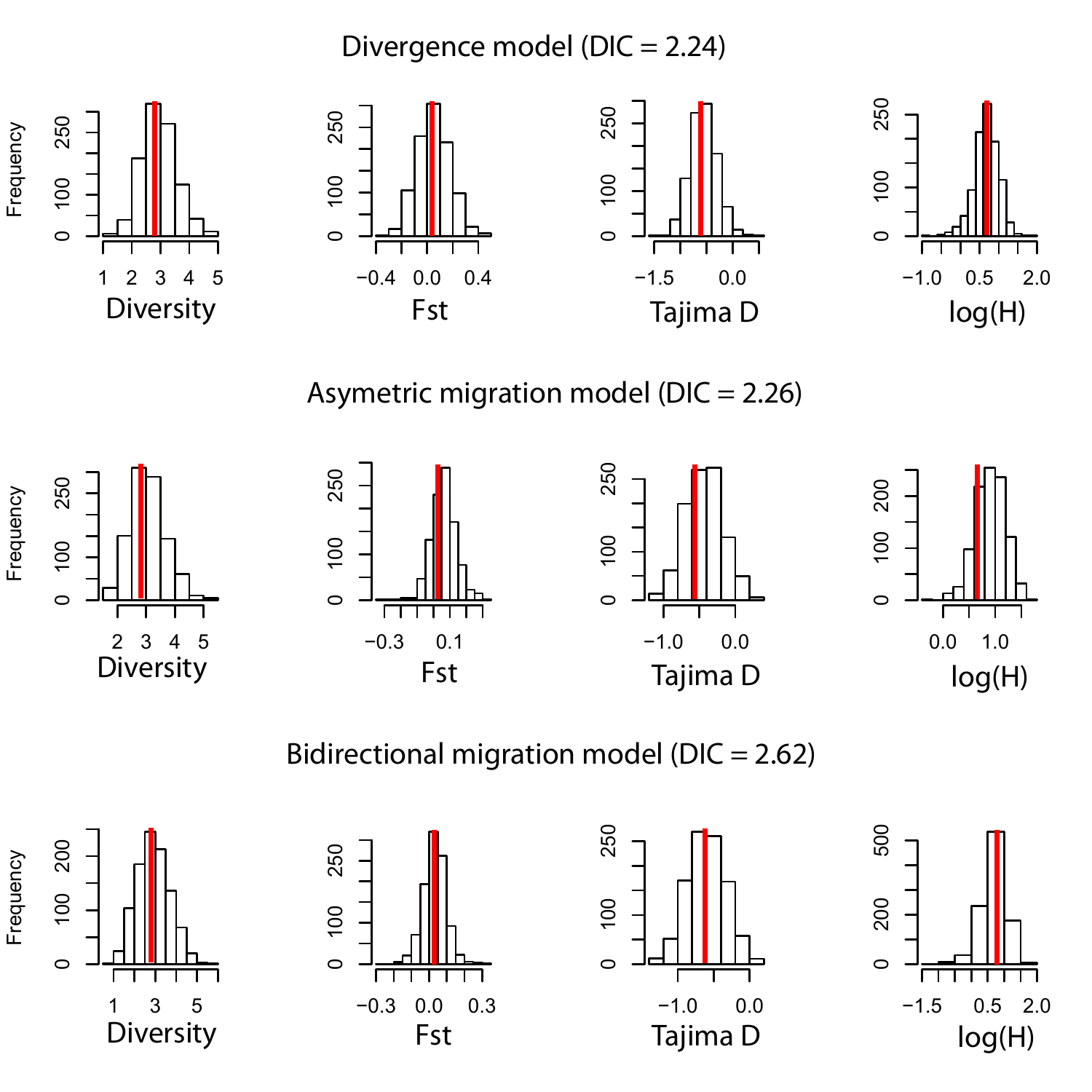}
\end{center}
\caption{\em Posterior predictive checks for three isolation with migration models. Data were simulated under the bidirectional migration model (third model). The summary statistics were within the tails of the posterior predictive distributions under all models. The vertical bars correspond to the observed summary statistics.}
\end{figure}

\section*{Human data analysis}

In this section, we used the deviance information criterion to question the replacement of Neanderthals by modern human using modern human DNA. Genetic data based on resequencing from noncoding regions were utilized to discriminate between models of divergence incorporating various levels of admixture between the two species (Laval et al 2010). We used sequence variation surveyed in DNA samples from 213 healthy donors. The panel included 118 sub-Saharan African individuals, 47 European individuals, and 48 East-Asian individuals. Following Laval et al (2010), we re-analyzed 20 autosomal regions (27 kb per individual, mean sequence length per region of 1.33 kb) that met criteria of selective neutrality. The regions were selected to be independent from each other, residing at least 200 kb apart from any known or predicted gene or spliced expressed sequence tag, and not to in linkage disequilibrium with any known or predicted gene. 

To run an ABC analysis, we used a coalescent-based algorithm implemented in SIMCOAL 2 (Laval et al 2004), we generated synthetic data that consisted of 20 independent DNA sequences of 1.4 kb each. The mutation and the recombination rates of each region were drawn from gamma distributions (Table 3). The evolutionary scenarios assumed an early diffusion of archaic hominids out of Africa $\sim$1.25 and $\sim$2.25 million years ago and an African exodus of modern humans between $\sim$40,000-100,000 years ago (Fagundes et al 2007). By tuning the replacement rate, $\delta$, we considered various levels of introgression of archaic genetic material into the modern human gene pool. Nineteen models differing in their prior distributions on $\delta$ were considered. In these models, each prior distribution was uniform over an interval of length 0.1, from $0<\delta <0.1$ to $0.9<\delta <1$. Each interval was deduced from the previous one by a translation of $h = 0.05$. Our objective was to discriminate among models with low, medium or high levels of admixture.    

Under equilibrium assumptions, the human effective population size has been estimated at $\sim$10,000 individuals on the basis of human-chimpanzee divergence and intra-specific linkage disequilibrium levels (Harpending et al 1998). To give population size a degree of freedom and to match with a consensus estimate of human populations, we defined a gamma prior distribution with a mean of 10,000 individuals and a .95 confidence interval of 3,000 to 21,000 individuals (Table 3). In all models  we considered a constant size for the African, Asian and European modern populations (Laval et al 2010).

For each genomic region, we specifically computed global and pairwise $F_{\rm ST}$, based on haplotype frequencies, the number of haplotypes, $K$, the number of polymorphisms, $S$, the nucleotide diversity, $\pi$, and Tajima's $D$, the expected heterozygosity. We also computed the variance between regions for $\pi$ and $D$.
The summary statistics were calculated by merging all population samples (except for population differentiation indices) in order to minimize the effects of recent demographic events related to the continental populations. 

\begin{table}
\begin{center}
\begin{tabular}{cccccc}
\hline
&&&&&\\
Parameter & & mean & min & max & prior\\
&&&&&\\
\hline
DNA parameters &&&&&\\
\hline
					   
	{\small Mutation rate} &	$\mu$ &	2.5$\times 10^{-8}$ &	1.3$\times10^{-8}$	& 5.05$\times10^{-8}$	& ~G \\	
   
	{\small Recombination rate}&	$\rho$	&$10^{-8}$	&$0.1\times 10^{-8}$	&$1.5 \times 10^{-8}$	&~G \\

Demographic parameters &&&&&\\	   
							   
	{\small Ancestral population size}	& $N'$	&10000&	500&	40000&	~G \\	   
	{\small Time of exodus from Africa}	& $T_E$ &	1.88$\times 10^{6}$	&1.25$\times 10^{6}$&	2.5$\times 10^{6}$ &	~U \\	   
	{\small Time of European-Asian split}	& $T_{E-EA}$	&25010&	12520&	37500&	~U \\	   
	{\small Migration rate}	 & m	&$ 2\times 10^{-4}$	&$10^{-6}$	&4$\times10^{-3}$&	ND \\
	   
	{\small African population size} &	$N_A$	&10000&	500	&40000&	~U\\	   
	{\small European population size}	&$N_E$&	10000&	500&	40000&	~U\\
	   
	{\small East-Asian population size}	&$N_{EA}$&	10000&	500	&40000&	~U	\\
   
\hline	
\hline
\end{tabular}
\end{center}
\caption{\em Description of the prior distributions of simulated parameters. U and G denote Uniformly and Gamma distributed distributions, ND (for not drawn). Times are expressed in number of years (generation times of 25 years). The  modern migration rates,  $m$, is the proportion of migrants after the Out-of-Africa exodus.  The mutation rate, $\mu$, is expressed in per generation per base, and  the recombination rate, $\rho$, is expressed in per generation per pair of adjacent bases.}
\end{table}

Considering 19 models with distinct prior distributions for the replacement rate of Neanderthals by modern humans, we found a decreasing trend  both in the expected deviance $\bar{D}_1$ and in DIC$_1$. For high values of the replacement rate, $\delta$, $\bar{D}_1$ and DIC$_1$ were close to each other (Figure 7). These indices favored models exhibiting high values of the parameter $\delta$ and low levels of genetic introgression of the human genetic pool by Neanderthal genes (Green et al 2010). 

\begin{figure}
\begin{center}
\includegraphics[width=9cm,angle=0]{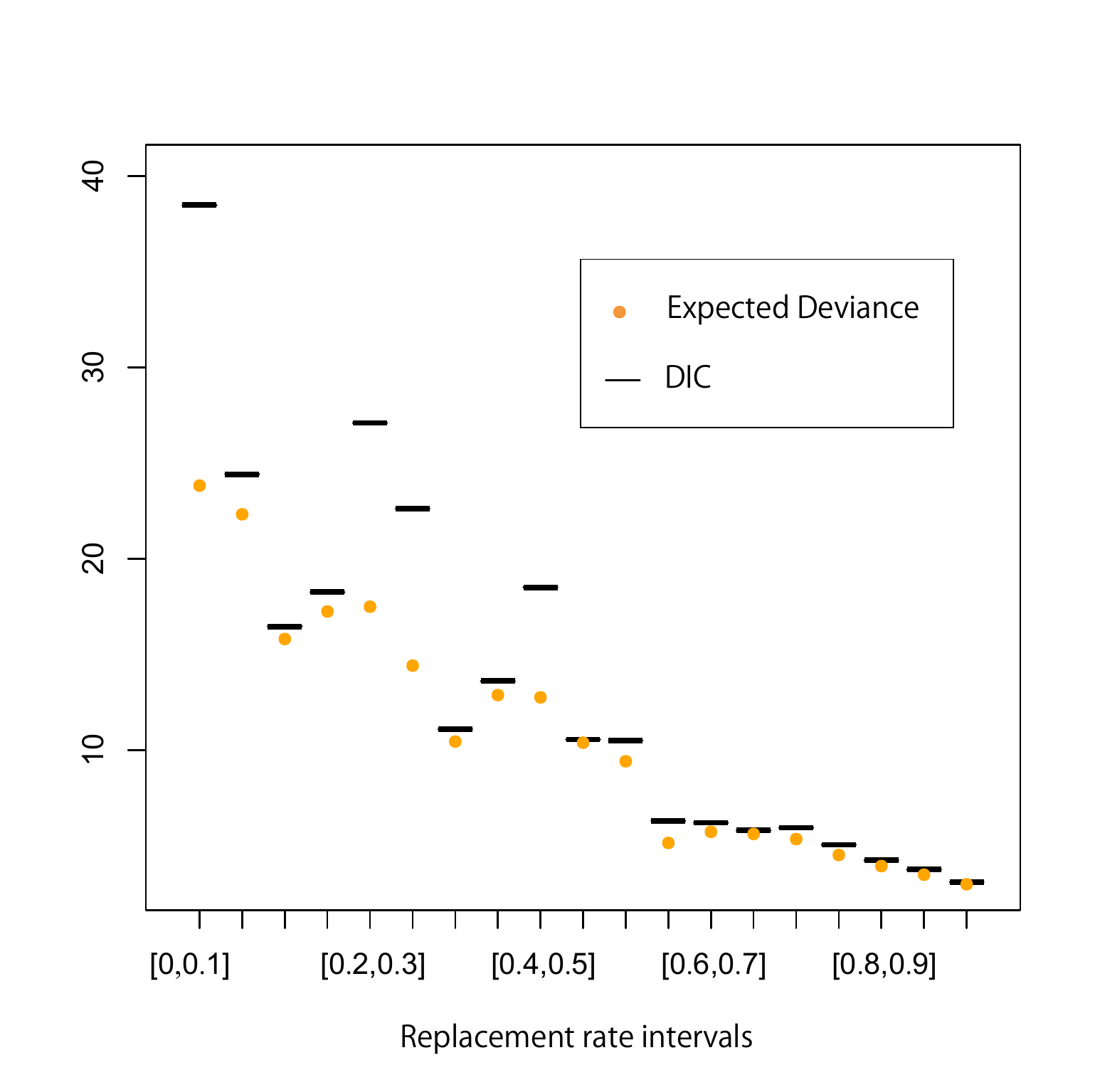}
\end{center}
\caption{\em Replacement of Neanderthals by modern humans. Expected Deviance, $\bar{D}_1$, and DIC$_1$ for models with different prior distributions on the replacement rate, $\delta$. The prior distributions are uniform over intervals of length 0.1. Each interval is deduced by translation of $h = 0.05$ from the previous one.}
\end{figure}

\section*{Conclusion}

Model selection using ABC algorithms is notoriously difficult (Csill\'ery et al 2010b), and it has been the topic of an intense debate in evolutionary genetics (Templeton 2009, Beaumont et al 2010, Robert et al 2011). As we have shown in this study, some approximations of model probabilities computed by counting replicates from each model falling at a distance smaller than a given value, $\epsilon$, to the observed data can potentially lead to systematic errors in ABC, especially when $\epsilon$ is not small. Using small acceptance rates implies, however, that gigantic numbers of simulations are performed, especially when more than ten summary statistics are used.

Our purpose here was not to argue that any model choice previously performed in ABC studies on the basis of approximate model probabilities is unreliable. Indeed, inconsistencies in model predictions may still be detected by using standard posterior predictive model checking procedures. In addition we do not argue that regression adjustments are problematic, and the results obtained under our simulation models should motivate users to apply these corrections systematically.

Regression adjustments can correct the posterior values obtained from the rejection algorithm, but they have no effect on posterior model probabilities. To overcome this potential issue, we argue that model selection should be done for the statistical distributions that correspond to the transformed model, and we propose an approach based on the evaluation of posterior predictive quantities. Our solution is based on the formulation of an approximate deviance, and  bypasses the estimation of posterior model probabilities. Our simulation study showed that the concepts of approximate deviance provide reasonable answers to the model choice issue in the population genetic examples tested, which are representative of this field (Beaumont 2010, Nielsen and Wakeley 2001). Finally we emphasize that since the computation of DIC is based on posterior predictive distributions, this approach applies to any type of ABC algorithm.

\bibliographystyle{empty}

\begin{thebibliography}{9}

\bibitem{akaik:74} Akaike H (1974) A new look at the statistical model identification. IEEE Trans Automat Control 19: 716 - 723.

\bibitem{bazin:10} Bazin E, Dawson KJ, Beaumont MA (2010) Likelihood-free inference of population structure and local adaptation in a Bayesian hierarchical model. Genetics 185: 587-602.

\bibitem{beaum:02} Beaumont MA, Zhang W, Balding DJ (2002) Approximate Bayesian computation in population genetics. Genetics 162: 2025 - 2035.

\bibitem{beaum:08} Beaumont MA (2008) Joint determination of topology, divergence time, and immigration in population trees. In Simulation, Genetics, and Human Prehistory (Matsumura, S., Forster, P. and Renfrew, C., eds) McDonald Institute for Archaeological Research.

\bibitem{beaum:09} Beaumont MA, Cornuet JM, Marin JM, Robert CP (2009) Adaptive approximate Bayesian computation. Biometrika 96: 983-990. 

\bibitem{beaum:10} Beaumont MA, Nielsen R, Robert CP, Hey J, Gaggiotti OE, et al (2010) In defence of model-based inference in
phylogeography. Molecular Ecology 19: 436 - 446.

\bibitem{beaum:11} Beaumont MA (2010) Approximate Bayesian Computation in Evolution and Ecology. Annual Review of Ecology, Evolution, and Systematics 41: 379-406.

\bibitem{borto:07} Bortot P, Coles SG, Sisson SA (2007) Inference for stereological extremes. J Am Stat Assoc 102: 84 - 92.

\bibitem{blum:10} Blum MGB, Fran\c cois O (2010) Non-linear regression models for Approximate Bayesian Computation. Statistics and Computing 20:63 - 73.

\bibitem{burnh:02} Burnham KP, DR Anderson (2002)  Model selection and multimodel inference: a practical information-theoretic approach. 2nd Edition. Springer-Verlag, New York, New York, USA. 

\bibitem{carst:09} Carstens BC, Stoute HN, Reid NM (2009) An information-theoretical approach to phylogeography. Molecular Ecology 18: 4270 - 4282

\bibitem{celeu:06} Celeux G, F Forbes, CP Robert, DM Titterington (2006) Deviance information criteria for missing data models (with Discussion). Bayesian Analysis 1: 651-706.

\bibitem{cornu:09} Cornuet JM, Santos F, Beaumont MA, Robert CP,
Marin JM, Balding DJ, Guillemaud T, Estoup A (2008) Inferring population history with DIY ABC: a user-friendly approach to approximate Bayesian computation. Bioinformatics 24:2713-2719. 

\bibitem{csill:10} Csill\'ery K, Blum MGB, Gaggiotti OE, Fran\c cois O (2010) Approximate Bayesian Computation in practice. Trends in Ecology and Evolution 25: 410-418.

\bibitem{csill:10b} Csill\'ery K, Blum MGB, Gaggiotti OE, Fran\c cois O (2010b) Invalid arguments against ABC: A reply to AR Templeton. Trends in Ecology and Evolution 25: 490-491.


\bibitem{fagun:07} Fagundes NJ, Ray N, Beaumont M, Neuenschwander S, Salzano FM, et al (2007) Statistical evaluation of alternative models of human evolution. Proc Natl Acad Sci USA 104: 17614 - 17619. 

\bibitem{fay:00} Fay JC, CI Wu (2000) Hitchhiking under positive Darwinian selection. Genetics 155: 1405 - 1413.

\bibitem{franc:08} Fran\c cois O, MGB Blum, M Jakobsson, N Rosenberg  (2008) Demographic history of European populations of {\it Arabidopsis thaliana}. PLoS Genetics 4: e1000075.

\bibitem{gelma:04} Gelman A, JB Carlin, HS Stern, DB Rubin (2004) Bayesian Data Analysis (2nd edn), Chapman and Hall, New York. 


\bibitem{green:10} Green R, Krause J, Briggs AW, Maricic T, Stenzel  T (2010) A draft sequence of the Neandertal genome. Science 328:710 - 722. 

\bibitem{grela:09} Grelaud A, CP Robert, JM Marin, F Rodolphe and JF Taly (2009) ABC likelihood-free methods for model choice
in Gibbs random fields. Bayesian Analysis 4: 317-336.

\bibitem{harpe:98} Harpending HC, Batzer MA, Gurven M, Jorde LB, Rogers AR, et al. (1998) Genetic traces of ancient demography. Proc Natl Acad Sci USA 95: 1961 - 1967.

\bibitem{hey:03} Hey J, Machado CA (2003) The study of structured populations: a difficult and divided science has new hope. Nature Reviews Genetics 4: 535-543. 

\bibitem{hudso:92} Hudson RR, Slatkin M, Maddison WP
(1992) Estimation of levels of gene flow from DNA sequence data.
Genetics 132: 583-589. 

\bibitem{hudso:02} Hudson RR (2002) Generating samples under a Wright-Fisher neutral model. Bioinformatics 18:337-338.

\bibitem{johns:04} Johnson JB, Omland KS (2004) Model selection in ecology and evolution. Trends in Ecology and Evolution 19: 101 - 108. 

\bibitem{jeffr:61} Jeffreys H (1961) The Theory of Probability. Oxford University Press, Oxford UK.

\bibitem{laval:04}  Laval G, Excoffier L (2004) SIMCOAL 2.0 A program to simulate genomic diversity over large recombining regions in a subdivided population with a complex history. Bioinformatics 20: 2485 - 2487.

\bibitem{laval:10} Laval G, Patin E, Barreiro LB, Quintana-Murci L (2010) Formulating a historical and demographic model of recent human evolution based on resequencing data from noncoding regions. PLoS ONE 5(4): e10284. 

\bibitem{leuen:09}  Leuenberger C, Wegmann D (2010) Bayesian computation and model selection without likelihoods. Genetics 184: 243-252.

\bibitem{lopes:09} Lopes J, Beaumont MA (2010) ABC: A useful Bayesian tool for the analysis of population data. Infection, Genetics and Evolution 10: 825-832.


\bibitem{marjo:03} Marjoram P, Molitor J, Plagnol V, Tavar\'e S (2003) Markov chain Monte Carlo without likelihoods. Proc Natl Acad Sci USA 100: 15324 - 15328.

\bibitem{marjo:06} Marjoram P, Tavar\'e S (2006) Modern computational approaches for analysing molecular genetic variation data. Nature Review Genetics 7: 759 - 770.


\bibitem{nielsen:01} Nielsen R, Wakeley J (2001) Distinguishing migration from isolation: A Markov chain Monte Carlo approach.
Genetics 2001 158: 885-896.


\bibitem{patin:09} Patin E, Laval G, Barreiro LB, Salas A, Semino O, et al. (2009) Inferring the demographic history of African farmers and Pygmy hunter – gatherers using a multilocus resequencing data set. PLoS Genet 5: e1000448.


\bibitem{pritc:99}  Pritchard JK, Seielstad MT, Perez-Lezaun A, Feldman MW (1999) Population growth of human Y chromosomes: a study of Y chromosome microsatellites. Mol Biol Evol 16: 1791 - 1798.

\bibitem{r:10} R Development Core Team (2010) R: A language and environment for statistical computing. R Foundation for Statistical Computing, Vienna, Austria.

\bibitem{ratma:09} Ratmann O, Andrieu C, Wiuf C, Richardson S (2009) Model criticism based on likelihood-free inference. Proc Natl Acad Sci USA 106: 10576-10581.

\bibitem{riple:04} Ripley BD (2004) Selecting amongst large classes of models. In Methods and Models in Statistics: In Honour of Professor John Nelder, FRS (Adams, N., Crowder, M, Hand, D.J. and Stephens, D., eds), pp. 155-170 Imperial College Press.


\bibitem{robert:11} Robert CP, Cornuet J-M, Marin J-M, Pillai NS (2011) Lack of confidence in approximate Bayesian computational (ABC) model choice. arXiv:1102.4432.

\bibitem{sisso:07} Sisson SA, Fan Y, Tanaka MM (2007) Sequential Monte Carlo without likelihoods. Proc Natl Acad Sci USA 104: 1760 - 1765.

\bibitem{spieg:02} Spiegelhalter DJ, Best NG, Carlin BP, van der Linde A (2002) Bayesian measures of model complexity and fit. J R Stat Soc Ser B 64: 583 - 639.


\bibitem{tavar:04} Tavar\'e S (2004) Ancestral inference in population genetics. In: École d'Été de Probabilités de Saint-Flour XXXI – 2001, Lecture Notes in Mathematics. Edited by Cantoni O, Tavar\'e S, Zeitouni O, eds) Berlin: Springer-Verlag.

\bibitem{tajim:89} Tajima F (1989) Statistical method for testing the neutral mutation hypothesis by DNA polymorphism. Genetics 123: 585 - 595.


\bibitem{templ:09} Templeton AR (2009) Statistical hypothesis testing in intraspecific phylogeography: NCPA versus ABC. Molecular Ecology 18: 319 - 331.

\bibitem{thorn:06} Thornton KR, Andolfatto P (2006) Approximate Bayesian inference reveals evidence for a recent, severe, bottleneck in a Netherlands population of {\it Drosophila melanogaster}. Genetics 172: 1607-1619.

\bibitem{thorn:09} Thornton KR (2009) Automating approximate Bayesian computation by local linear regression. BMC Genetics 10:35.

\bibitem{toni:09}  Toni T, Welch D, Strelkowa N, Ipsen A, Stumpf MP (2009) Approximate Bayesian computation scheme for parameter inference and model selection in dynamical systems. J R Soc Interface 6: 187 - 202.

\bibitem{toni:10} Toni T, Stumpf MP (2010) Simulation-based model selection for dynamical systems in systems and population biology. Bioinformatics 26:104-110. 

\bibitem{wegma:09} Wegmann D, Leuenberger C, Excoffier L (2009) Efficient approximate Bayesian computation coupled with Markov Chain Monte Carlo without likelihood. Genetics 182: 1207-1218.

\bibitem{wilki:09} Wilkinson R (2008) Approximate Bayesian computation (ABC) gives exact results under the assumption of model error. arXiv:0811.3355v1 [stat.CO].

\end{thebibliography}

\end{document}